\documentclass[english]{article}
\usepackage[T1]{fontenc}
\usepackage[latin9]{inputenc}
\usepackage{amssymb}
\usepackage{esint}
\usepackage{babel}
\begin{document}

\title{{\large{}Equations for generalized n-point information with extreme
and not extreme approximations in the free Fock space }}

\author{{\normalsize{}Jerzy Hanckowiak}}

\author{{\normalsize{}(Former lecturer of Wroclaw and Zielona Gora Universities)}}

\author{{\normalsize{}e-mail: jerzy.hanckowiak@gmail.com}}

\author{{\normalsize{}Poland, EU}}

\date{{\normalsize{}7. November. 2015 }}
\maketitle
\begin{abstract}
{\normalsize{}The general n-point information (n-pi) are introduced
and equations for them are considered. The role of right and left
invertible interaction operators occurring in these equations together
with their interpretation is dicussed. Some comments on aproximations
to the proposed equations are given. The importance of positivity
conditions and a possible interpretation of n-pi in the case of their
non-compliance, for essentially nonlinear interactions (ENI), are
proposed. A language of creation, annihilation and projection operators
which can be applied in classical as well as in quantum case is used.
The role of the complex numbers and functions in physics is also a
little elucidated.}{\normalsize \par}

{\normalsize{}\tableofcontents{}}{\normalsize \par}
\end{abstract}

\section{{\large{}Introduction}}

In many cases, whether due to too sensitivity of the systems to small
changes of initial and/or boundary conditions or its complexity, are
not important their detailed description. In such cases we are using
averages and higher and higher correlations among averages, which
we called the n-pi (n-point information), see Eq.\ref{eq:2}. 

For the first time, the general n-pi with multiple-times instead of
one time were introduced by Kraichnan and Lewis in \cite{Kraich 1962}
in order to describe in a more complete way the random properties
of the system S characterized by the 'field' $\varphi[\tilde{x};\alpha]$.
Having similar intensions, we generalize n-pi even more through introduction
multiple-initial times and multiple-additional (initial and/or boundary)
conditions. In other words, the \textbf{tensor products} of the fields
concern not only space time variables but also additional conditions
imposed on the fields, see \ref{eq:6-1}. It is interesting that \textbf{all
these generalizations satisfy the same equations}, \ref{eq:15}, and
hence - similar formulas can be derived with their help! 

In the paper, we discuss again, from the point of view of an expected
new and more fundamental theory, certain problems concerning a role
and inerpretation of the left and right invertible operators, Sec.4
. In Secs 4 and 5 we consider perturbation expansions with the two
canonical extreme assumptions concerning zeroth order approximations.
In Sec.7 we avoid such extreme assumptions. In Sec.8 we give some
remarks concerning operators describing essentially nonlinear (ENI)
interactions. Some general remarks are given in Sec 9-10. In Sec.11
an example of n-pi satisfying the positivity conditions is given. 

We also show how - by the relevant averaging - a role of the equations
describing in detail the system is gradually diminished to eventually
replace them by the energy integral, Sec.12,. Below, in subsections,
we put the two basic conditions that must be satisfied by the n-pi.

To avoid excess of indexes, $\tilde{x}$ contains not only the space-time
variables (space-time localization) including the initial time $t^{0}$but
also subscripts and superscripts indexes denoting components of the
field $\varphi$. In other words, $\tilde{x}$ may describe the \textit{external
and internal particle or field variables}. So we have:

\begin{equation}
\tilde{x}=(\vec{x},t,t^{0},_{\mu,\nu,\xi,...,}^{\eta,\theta,\vartheta,...,},)\in R^{\tilde{d}}\label{eq:1}
\end{equation}
 It means that the field $\varphi$ is defined (lives) at the larger
space which rather should be called a set in which the addition operation
is not defined for all components of the 'vector' $\tilde{x}$. The
usual space-time space of physics is denoted by $R^{d}\ni(\vec{x},t)$
or $M^{d}$. In fact we assume a discrete versions of these spaces,
hence, in this and other papers, we consider and have been considered
the \textbf{difference equations}. The initial and/or boundary conditions
are denoted by $\alpha$. For the n point particles system, we usually
use another notation and then

\begin{equation}
\varphi\equiv q,\: and\:\tilde{x}=(t,i,j)\label{eq:2-2}
\end{equation}
where $t$ is the time, $i=1,...,n$ is numbering the individual particles
and $j=1,2,3$ refers to a particle components. This case is called
somewhat artificially the \textit{zero-dimentional (0-D) field}. In
fact, by assigning to different components and their values of the
vector $\tilde{x}$ - the corresponding components of the physical
fields, one can consider only the one superfield $\varphi$ describing
the whole physical system. 

The field $\varphi$ at points $\tilde{x}$ may have no physical meaning
if it is very sensitive to even small changes of $\alpha$. Then,
physical meaning may have some averaged or smoothed values of the
field $\varphi$ with respect to $\alpha$ or some components of $\tilde{x}$,
which in all cases are denoted by $<\varphi(\tilde{x})>$. However,
the problem with averaged quantities is such that equations for them
are not closed and to get a complete system of equations, we are forced
to introduce other quantities than averaged or smooth solutions of
the starting equations which are denoted as $<\varphi(\tilde{x}_{1})...\varphi(\tilde{x}_{n})>$,
$n=1,\cdots,\infty$ and which are called the n-pi whose particular
cases are correlation functions. The symbol $<...>$means in fact
an \textit{averaging} or \textit{a smoothing} of products $\varphi[\tilde{x}_{1};\alpha]\cdots\varphi[\tilde{x}_{n};\alpha]$
with respect to the variables or functions $\alpha$. For example,
in the case of averaging

\begin{equation}
<\varphi(\tilde{x}_{1})...\varphi(\tilde{x}_{n})>=\int\delta\alpha\left\{ \varphi[\tilde{x}_{1};\alpha]\cdots\varphi[\tilde{x}_{n};\alpha]\right\} P[\alpha]\label{eq:2}
\end{equation}
 where $P[\alpha]$ is a probability density, but in the case of smoothing,
$P[\alpha]$ is some weight function or functional. $\int\delta\alpha$
means the formal functional integration or ordinary integrations or
a sum $\sum$ in the case of discrete variables.

\subsection{Positivity conditions}

The positivity of the $P[\alpha]\geqq0$ leads to the \textit{positivity
conditions} upon the even n-pi:

\begin{equation}
<\varphi(\tilde{x}_{1})...\varphi(\tilde{x}_{2n})>\geqq0\label{eq:2-1}
\end{equation}
if for every $\tilde{x}_{i}$ there is at least one $\tilde{x}_{j}$
with $j\neq i$ such that $\tilde{x}_{i}=\tilde{x}_{j}$, see , eg.,
one of the author's papers. This is the \textit{positivity principle
}which expresses integral structure \ref{eq:2} of the n-pi.

\subsection{{\normalsize{}P}ermutation symmetry condition}

From \ref{eq:2} it is seen that these n-pi are permutation symmetric: 

\begin{equation}
V(\tilde{x}_{(n)})\equiv<\varphi(\tilde{x}_{1})...\varphi(\tilde{x}_{n})>=<\varphi(\tilde{x}_{i_{1}})...\varphi(\tilde{x}_{i_{n}})>\label{eq:3}
\end{equation}
where $(i_{1},...,i_{n})$ denotes arbitrary permutation of the number
sequence $(1,...,n)$ and $\tilde{x}_{i}\in R^{\tilde{d}}$. The above
symmetry takes place 'iff' we are exchanging among themselves all
components of 'vectors', e.g, $\tilde{x}_{i}$and $\tilde{x}_{j}$.
However, if you do not exchange at least one pair of the components,
e.g, $t_{i}^{0}$and $t_{j}^{0}$ , then the n-pi may not be symmetric
with respect to permutation of the space-time variables and at this
point even the classical n-pi are not permutatin symmetrical.

From formula \ref{eq:2} it is also directly seen that any approximation
to the field $\varphi[\tilde{x};\alpha]$ preserves the positivity
conditions and the permutation symmetry of the approximated n-pi. 

One can get even more general n-pi, if every $\alpha$ describing
initial and/ or boundary conditions is the sum of two terms:

\begin{equation}
\alpha=\beta+\gamma\label{eq:5-1}
\end{equation}
 where $\beta$ can describe different initial and boudary conditions
in every field $\varphi[\tilde{x};\alpha]$ entering the n-point smoothing
integral:

\begin{equation}
<\varphi(\tilde{x}_{1})...\varphi(\tilde{x}_{n});\beta_{(n)}>=\int\delta\gamma\left\{ \varphi[\tilde{x}_{1};\beta_{1}+\gamma]\cdots\varphi[\tilde{x}_{n};\beta_{n}+\gamma]\right\} P[\gamma]\label{eq:6-1}
\end{equation}
 In the case that all $\beta_{i}$ will not be equal to each other,
the permutation symmetry \ref{eq:3} and positivity conditions \ref{eq:2-1}
are not satisfied in general. 

The structure of n-pi, \ref{eq:6-1}, shows also surprising property
of n-pi. If some systems do not interacting with one another, and
in spite of that we want to describe it as a one system, then the
corresponding n-pi do not need to be the products of corresponding
n'-pi and n''-pi, (n'+n''=n), where n' is related to one system and
n'' is related to another system, Sec.11. All depends on the weight
$P[\gamma]$ in the above averaging integral. From the point of view
of n-pi, it looks as if the knowledge of the systems included in the
$P[\gamma]$ , for example, that:

\begin{equation}
P[\gamma]\neq P[\gamma']\star P[\gamma'']\label{eq:8-1}
\end{equation}
imitates an interaction between not interacting systems!!?

\section{{\large{}Generating vectors for n-pi. Cuntz relations and the unit
operator. Free Fock space (FFS), }}

In the Dirac notationa, a generating vector $|V>$, for the infinite
collection of n-pi $V(\tilde{x}_{(n)})$, is defined as follows:

\begin{equation}
|V>=|0>+\sum_{n=1}^{\infty}\int d\tilde{x}_{(n)}V(\tilde{x}_{(n)})|\tilde{x}_{(n)}>\label{eq:4}
\end{equation}
 where the orthonormal base vectors $|\tilde{x}_{(n)}>$ are introduce.
Here $\tilde{x}_{(n)}\equiv\tilde{x}_{1},...,\tilde{x}_{n}$ and so
on. They saytisfy the following relations

\begin{equation}
<\tilde{y}_{m},...,\tilde{y}_{1}|\tilde{x}_{1},...,\tilde{x}_{n}>=\delta_{mn}\delta(\tilde{y}_{1}-\tilde{x}_{1})\cdots\delta(\tilde{y}_{n}-\tilde{x}_{n})\label{eq:5}
\end{equation}
where $m,n=0,1,2...$, and $\delta_{mn}$is the Kronecker delta, and
$\delta(\tilde{y}_{1}-\tilde{x}_{1})\cdots\delta(\tilde{y}_{n}-\tilde{x}_{n})$
are products of the Dirac and Kronecker delta functions or only Kronecker
functions if all components of 'vectors' $\tilde{y},\tilde{x}$ are
discrete. $|\tilde{x}_{0}>\equiv|0>$. 

In contrast to the canonical, classical n-pi V, the base vectors are
not permutation symmetric, nevertheless we get: 

\begin{equation}
<\tilde{y}_{(n)}|V>=V(\tilde{y}_{(n)})\label{eq:6}
\end{equation}
 also for permutation symmetric n-pi $V$. 

If such base vectors are constructed by means of the 'creation' and
'annihilation' operators (Here and below the Asian hat over the letter
indicates that the letter represents an operator in the space under
consideration)

\begin{equation}
|\tilde{x}_{(n)}>=\hat{\eta}^{\star}(\tilde{x}_{1})\cdots\hat{\eta}^{\star}(\tilde{x}_{n})|0>\label{eq:7}
\end{equation}
 and

\textsuperscript{}

\begin{equation}
<\tilde{x}_{(n)}|=<0|\hat{\eta}(\tilde{x}_{n})\cdots\hat{\eta}(\tilde{x}_{1})\label{eq:8}
\end{equation}
which satsfy the Cuntz relations:

\begin{equation}
\hat{\eta}(\tilde{x})\hat{\eta}^{\star}(\tilde{y})=\delta(\tilde{x}-\tilde{y})\hat{I}\label{eq:9}
\end{equation}
 and additional restrictions are imposed:

\begin{equation}
\hat{\eta}(\tilde{x})|0>=0,\;<0|\hat{\eta}^{\star}(\tilde{y})=0\label{eq:10}
\end{equation}
 where the star over the operator means a linear operation, called
also involution, with the following properties: 

\begin{equation}
\left(\hat{\eta}^{\star}(\tilde{y})\right)^{\star}=\hat{\eta}(\tilde{y})\label{eq:11}
\end{equation}

\begin{equation}
\left(\hat{\eta}(\tilde{x}\hat{\eta}(\tilde{y})\right)^{\star}=\hat{\eta}^{\star}(\tilde{y})\hat{\eta}^{\star}(\tilde{x})\label{eq:12}
\end{equation}
then the space formed by the vectors \ref{eq:4} is called the \textit{free
Fock space} (FFS). In fact, the notions of creation and annihilation
operators were also used by Dirac, see \cite{Farm 2009}, who was
assuming commutation or antycommutation properties for the creation
operators among themselvs. The Cuntz operators, without these properties
but only with: \ref{eq:9} and \ref{eq:10}, allows to use the notion
of creation and annihilation operators in classical as well as in
quantum physics and the corresponding Fock space (FS) is called the
free Fock space (FFS). An equivalent description of the FFS without
using the creation and annihilation operators is given by author in
\cite{Han 1986} by means of the linear functionals. 

In the FFS, the unit operator can be expressed as follows

\begin{equation}
\hat{I}=|0><0|+\int d\tilde{x}\hat{\eta}^{\star}(\tilde{x})\hat{\eta}(\tilde{x})\label{eq:13}
\end{equation}
 Remember, $\int d\tilde{x}$ is a summation and integration, or summation
only as we usually assume.

\section{{\large{}Equations for n-pi and their structure}}

With the help of unique 'field' $\varphi$ (discrete indexes are also
contained in the symbol $\tilde{x}$), we consider the following integro-differential
or rather the \textbf{difference equations}: 

\begin{equation}
L[\tilde{x};\varphi]+\lambda N[\tilde{x};\varphi]+G(\tilde{x})=0\label{eq:14}
\end{equation}
 for the field $\varphi$ which can describe an electromagnetic or
hydromechanic or other equations of motion. Here the terms L and N
describe a linear and nonlinear dependence on the field $\varphi$,
the term $G$ does not depend on $\varphi$ and can describe a source
of the field $\varphi$ or an external force. With the help of Eq.\ref{eq:14}
and definition of n-pi: \ref{eq:2},\ref{eq:3}, one can derive, see
,e.g., \cite{Han 2010}, the following equation for the generating
vector |V>:

\begin{equation}
\left(\hat{L}+\lambda\hat{N}+\hat{G}\right)|V>=\hat{P}_{0}\hat{L}|V>+\lambda\hat{P}_{0}\hat{N}|V>\equiv|0>_{info}\label{eq:15}
\end{equation}
 with the following operators:

\begin{equation}
\hat{L}=\int\hat{\eta}^{\star}(\tilde{x})L[\tilde{x};\hat{\eta}]d\tilde{x}+\hat{P}_{0},\label{eq:16}
\end{equation}

\begin{equation}
\hat{N}=\int\hat{\eta}^{\star}(\tilde{x})N[\tilde{x};\hat{\eta}]d\tilde{x}+\hat{P}_{0},\label{eq:17}
\end{equation}
 
\begin{equation}
\hat{G}=\int\hat{\eta}^{\star}(\tilde{x})G(\tilde{x})d\tilde{x}\label{eq:18}
\end{equation}
 where the projector

\begin{equation}
\hat{P}_{0}=|0><0|\label{eq:19}
\end{equation}
 $\lambda$ is is a theory parameter which can describe the strenth
of interaction reffered to as the coupling constant. We remind you
that $\int$ means integration and summation or, in the case of all
discrete variables - only summation. If L and N are given by the Volterra
series, then

\begin{equation}
\hat{L}[\tilde{x};\hat{\eta}]=\int d\tilde{y}L(\tilde{x};\tilde{x}_{1})\hat{\eta}(\tilde{x}_{1})\label{eq:20}
\end{equation}
 and

\begin{equation}
\hat{N}[\hat{x};\hat{\eta}]=\sum_{n=2}\int d\tilde{x}_{(n)}N(\tilde{x};\tilde{x}_{(n)})\hat{\eta}(\tilde{x}_{1})\cdots\hat{\eta}(\tilde{x}_{n})\label{eq:21}
\end{equation}
 For the \textit{local interaction}, for example:

\begin{equation}
N(\tilde{x};\tilde{x}_{(n)})=\delta(\tilde{x}-\tilde{x}_{1})\cdots\delta(\tilde{x}-\tilde{x}_{n})\label{eq:22}
\end{equation}

In the case of quantum fields $\varphi$, when the field $\varphi$
appearing in Eq.\ref{eq:14} is an operator, the n-pi as Wightman
functions satisfy identical equations as Eq.\ref{eq:15}. This means
that the \textbf{operator properties of the field $\varphi$ and appropriate
definition of the n-pi} must be only used in the zero approximation
to the considered equations for n-pi. (For different transformations
of Eq.\ref{eq:15}, see below). In other words, the FFS is a space
in which quantum and statistical properties of a system can be described
by the same equations. 

In the case of Green's n-pi it is possible another relation between
fields and operators $\hat{L},\hat{N},\hat{G}$ in which $\hat{G}\equiv\hat{C}$
is still lower triangular operator, but it has other projective properties,
see \cite{han 2011}. However, the remaining operators: $\hat{L}$and
$\hat{N}$ have similar structure in classical and quantum cases.

\section{{\large{}Right and left invertible operators, physical interpretation,
and transformed Eq.\ref{eq:15} }}

For a definition and properties of right and left invertible operators,
see Danuta Przeworska-Rolewicz book: ``Introduction to Algebraic
Analysis'' (1988), see also author's papers.

\subsection{Right invertible self-interaction and interaction operators}

Let us take the operator $\hat{L}\equiv\hat{L}_{r}$ related to the
linear part of Eq.\ref{eq:14}

\begin{equation}
\hat{L}_{r}=\int\hat{\eta}^{\star}(\tilde{y})L_{r}(\tilde{y},\tilde{x})\hat{\eta}(\tilde{x})d\tilde{y}d\tilde{x}+\hat{P}_{0}\label{eq:24}
\end{equation}
 with a function $L_{r}$ that there is such a function $L_{r}^{-1}$
that 

\begin{equation}
\int L_{r}(\tilde{x},\tilde{z})L_{r}^{-1}(\tilde{z},\tilde{y})d\tilde{z}=\delta(\tilde{x}-\tilde{y})\label{eq:25}
\end{equation}
 We define the operator 

\begin{equation}
\hat{L}_{r}^{-1}=\int\hat{\eta}^{\star}(\tilde{x})L_{r}(\tilde{x},\tilde{y})\hat{\eta}(\tilde{y})d\tilde{x}d\tilde{y}+\hat{P}_{0}\label{eq:26}
\end{equation}
 Then we get

\begin{equation}
\hat{L}_{r}\hat{L}_{r}^{-1}=\int d\tilde{y}\hat{\eta}^{\star}(\tilde{y})\hat{\eta}(\tilde{y})+\hat{P}_{0}=\hat{I}\label{eq:27}
\end{equation}
 see Sec.2. This means that $\hat{L}_{r}$ is the \textit{right invertible
operator} in the FFS and $\hat{L}_{r}^{-1}$ is a \textit{right inverse
operato}r. 

For a special important case of 

\begin{equation}
L_{r}(\tilde{x},\tilde{z})=\delta(\tilde{x}'-\tilde{z}')L_{r}(\tilde{x}'',\tilde{z}'')\label{eq:28}
\end{equation}
where, e.g., $\tilde{x}'+\tilde{x}''=\tilde{x}$ , and $\tilde{x}'\cdot\tilde{x}''=0$,
then

\begin{equation}
\hat{L}_{r}=\int\hat{\eta}^{\star}(\tilde{x})\delta(\tilde{x}'-\tilde{y}')(L_{r}(\tilde{x}'',\tilde{y}'')\hat{\eta}(\tilde{y})d\tilde{x}d\tilde{y}+\hat{P}_{0}\label{eq:29}
\end{equation}
if 

\begin{equation}
\int L_{r}(\tilde{x}'',\tilde{z}'')L_{r}^{-1}(\tilde{z}'',\tilde{y}'')d\tilde{z}''=\delta(\tilde{x}''-\tilde{y}'')\label{eq:30}
\end{equation}

Now, let take us an example of the \textit{local operator} $\hat{N}$
related to the nonlinear part of Eq.\ref{eq:14} describing \textit{self-interaction}
of the system elements :

\begin{equation}
\hat{N}\equiv\hat{N}_{r(self)}=\int d\tilde{y}\hat{\eta}^{\star}(\tilde{y})\hat{\eta}^{2}(\tilde{y})+\hat{P}_{0}\label{eq:31}
\end{equation}
 For discrete $\tilde{y}$, it is easy to show, Sec.2, that a right
inverse to $\hat{N}$ is

\begin{equation}
\hat{N}_{r(self)}^{-1}=\int d\tilde{y}\hat{\eta}^{\star}(\tilde{y})^{2}\hat{\eta}(\tilde{y})+\hat{P}_{0}\label{eq:32}
\end{equation}
 An example of the right invertible operator describing an interaction
of system elements is the operator

\begin{equation}
\hat{N}_{r(int)}=\int d\tilde{y}\hat{\eta}^{\star}(\tilde{y})\hat{\eta}(\tilde{y})\hat{\eta}(\tilde{y}+\tilde{a})+\hat{P}_{0}\label{eq:36-1}
\end{equation}
 where for the \textbf{\textit{local interaction}} the components
of vector $\tilde{a}$ related to the space time variables: $(\vec{a},t_{a})\equiv0$.
Some components of the vector $\tilde{a}$ related to identification
of the system elements \textbf{should be} different from zero. A right
inverse for this operator is:

\begin{equation}
\hat{N}_{r(int)}^{-1}=\int d\tilde{y}'\hat{\eta}^{\star}(\tilde{y}'+\tilde{a})\hat{\eta}^{\star}(\tilde{y}')\hat{\eta}(\tilde{y}')+\hat{P}_{0}\label{eq:37-1}
\end{equation}

At this point, and we will continue to assume that the indexes identifying
the individual elements of the system under consideration run an infinite
set of values, and variables, which do not correspond to any elements
we put identically equal to zero.

\subsection{Left invertible self-interaction operators }

It turns out that the local operators $\hat{N}$ can also be left
invertible. As an example let take a formal operator

\begin{equation}
\hat{N}\equiv\hat{N}_{l(self)}(\lambda')=\int d\tilde{x}\hat{\eta}^{\star}(\tilde{x})\frac{H(\tilde{x})\hat{I}}{\hat{I}-\lambda'\hat{\eta}(\tilde{x})}+\hat{P}_{0}\label{eq:33}
\end{equation}
 where $\lambda'$ is a new coupling constant usually called the \textit{minor
coupling constant} and $H$ is an arbitrary auxiliary function. In
particular, it can be equal to one. A left inverse to this operator,
denoted by $\hat{N}_{l}^{-1}$ and satisfying equation:

\begin{equation}
\hat{N}_{l(self)}^{-1}\hat{N}_{l(self)}=\hat{I}\label{eq:33-1}
\end{equation}
is

\begin{equation}
\hat{N}_{l(self)}^{-1}(\lambda')=\int d\tilde{y}E(\tilde{y})\left(\hat{I}-\lambda'\hat{\eta}(\tilde{y})\right)\hat{\eta}(\tilde{y})+\hat{P}_{0}\label{eq:34}
\end{equation}
 with 

\begin{equation}
\int d\tilde{x}E(\tilde{x})H(\tilde{x})=1\label{eq:35}
\end{equation}
 Other examples of the left invertible self-interaction operators
are 

\begin{equation}
\hat{N}_{l(self)}(\lambda')=\int d\tilde{x}\hat{\eta}^{\star}(\tilde{x})\frac{H(\tilde{x})\hat{I}}{\hat{I}-\lambda'\hat{\eta}^{m}(\tilde{x})}+\hat{P}_{0}\label{eq:39-2}
\end{equation}
where e.g.,$m\in R'\subset R$ , see also \cite{Han 2011'}. They
can be used for modeling the system S with self-interaction of its
0-D elements, which interact between each other in a linear way described
by the functional $L$, see Eq.\ref{eq:14}. It worth noting that
the above nonlinear term (the linear operator $\hat{N}$ in FFS depending
on the operator $\hat{\eta}$ in a nonlinear way) can be related to
an external field acting on the system S. We assume that a kinematic
term of the system S is also contained in the operator $\hat{L}$.
In this case the ``external'' interaction related to the operator
$\hat{G}$ depends only on the time $t$ and indexes $i,j$, see \ref{eq:2-2}.

\subsection{Left invertible, essentially nonlinear interaction (ENI) and self-interaction
operators (ENSI)}

The next step is to construct the left invertible operators in the
case of nonlinear interactions (not self-interaction). A simple modification
of the formula \ref{eq:39-2}:

\begin{equation}
\hat{N}_{l(int)}(\lambda')=\int d\tilde{x}\hat{\eta}^{\star}(\tilde{x})\frac{H(\tilde{x})\hat{I}}{\hat{I}-\lambda'\hat{\eta}(\tilde{x}+\tilde{a}_{1})\cdots\hat{\eta}(\tilde{x}+\tilde{a}_{m})}+\hat{P}_{0}=\hat{I}+\lambda'\hat{N}+\cdots+\hat{P}_{0}\label{eq:41-2}
\end{equation}
 where 'vectors' $\tilde{a}_{i}$ are defined as $\tilde{a}$ in subsection
4.2. In this case

\begin{equation}
\hat{N}_{l(int)}^{-1}(\lambda')=\int d\tilde{y}E(\tilde{y})\left(\hat{I}-\lambda'\hat{\eta}(\tilde{y}+\tilde{a}_{1})\cdots\hat{\eta}(\tilde{y}+\tilde{a}_{m})\right)\hat{\eta}(\tilde{y})+\hat{P}_{0}\label{eq:44-1}
\end{equation}

One can also use the left invertible formal operator:

\begin{equation}
\hat{N}_{l}(\lambda')=\hat{N}_{l}\frac{\hat{I}}{\hat{I}-\lambda'\hat{N_{r}}}\label{eq:45-1}
\end{equation}
 with a left invertible $\hat{N}_{l}$ and 'any' operator $\hat{N}_{r}$.
A left inverse to the operator $\hat{N}_{l}(\lambda')$ is

\begin{equation}
\hat{N}_{l}^{-1}(\lambda')=\left(\hat{I}-\lambda'\hat{N}_{r}\right)\hat{N}_{l}^{-1}\label{eq:46-1}
\end{equation}
 with a left inverse $\hat{N}_{l}^{-1}$ to the operator $\hat{N}_{l}$.
It is easy to check that the operator 

\begin{equation}
\hat{N}'_{l}(\lambda')=\frac{\hat{I}}{\hat{I}-\lambda'\hat{N}_{r}}\hat{N}_{l}\label{eq:47-1}
\end{equation}
 is also a left invertible operator with left inverse

\begin{equation}
\hat{N'}_{l}^{-1}(\lambda')=\hat{N}_{l}^{-1}\left(\hat{I}-\lambda'\hat{N}_{r}\right)\label{eq:48-1}
\end{equation}
 If all $\tilde{a}_{i}=0$ in the formula \ref{eq:41-2}, then the
operator $\hat{N}_{l(int)}(\lambda')$ desribes self-interaction.

\subsection{Transformed equations \ref{eq:15}{\normalsize{} }}

Let us assume first that in Eq.\ref{eq:15} the $\hat{L}\equiv\hat{L}_{r}$
is a right invertible operator with its right inverse $\hat{L}_{r}^{-1}$.
Multiplying Eq.\ref{eq:15} by $\hat{L}_{r}^{-1}$ we get an equivalent
equation:

\begin{equation}
\left(\hat{I}+\hat{L}_{r}^{-1}\hat{G}+\lambda\hat{L}_{r}^{-1}\hat{N}\right)|V>=\hat{L}_{r}^{-1}|0>_{info}+\hat{P}_{L_{r}}|V>\label{eq:36}
\end{equation}
 with any projection $\hat{P}_{L_{r}}|V>$ where the projector $\hat{P}_{L_{r}}=\hat{I}-\hat{L}_{r}^{-1}\hat{L}$. 

Now let us assume that the operator $\hat{N}$ in Eq.\ref{eq:15}
is a right invertible operator with its right inverse $\hat{N}_{r}^{-1}$.
We get similar equation to Eq.\ref{eq:36}:

\begin{equation}
\left(\lambda^{-1}\hat{N}_{r}^{-1}(\hat{L}+\hat{G})+\hat{I}\right)|V>=\lambda^{-1}\hat{N}_{r}^{-1}|0>_{info}+\hat{P}_{N_{r}}|V>\label{eq:37}
\end{equation}
 with the projector $\hat{P}_{N_{r}}=\hat{I}-\hat{N}_{r}^{-1}\hat{N}$.
This equation with any projection $\hat{P}_{N_{r}}|V>$ is also equivalent
to Eq.\ref{eq:15}. Now the problem arises how to choose vectors $\hat{P}_{L_{r}}|V>$and
$\hat{P}_{N_{r}}|V>$, which for the point of view Eq.\ref{eq:15}
may be any of the vectors of the respective subspace! In the perturbation
theory we solve this problem by choosing zero approximations in such
a way as \textbf{to satisfy the symmetry conditions, \ref{eq:44},
and the positivity principle, \ref{eq:2-1}}. See also a symmetrized
version of the equations as above. Then, at least, for small values
of parameters $|\lambda|$ or $|1/\lambda|$, one can fulfill the
above conditions. In all these considerations we assume that vectors
$\hat{L}_{r}^{-1}|0>_{info},\:\lambda^{-1}\hat{N}_{r}^{-1}|0>_{info}$
are trivial and are

\begin{equation}
\sim|0>\label{eq:52}
\end{equation}

Let us assume that the operator $\hat{N}$ is a left invertible operator
with the left invers operator $\hat{N}_{l}^{-1}$. Then, by multiplying
Eq.\ref{eq:15} by this left inverse operator, we get 

\begin{equation}
\left(\lambda^{-1}\hat{N}_{l}^{-1}(\hat{L}+\hat{G})+\hat{I}\right)|V>=\lambda^{-1}\hat{N}_{l}^{-1}|0>_{info}\label{eq:39}
\end{equation}
 It is also easy to see that in this case the operator 

\begin{equation}
\hat{P}_{N_{l}}=\hat{I}-\hat{N}\hat{N}_{l}^{-1}\label{eq:54-1}
\end{equation}
 is a projector. 

We do not consider the case of left invertible $\hat{L}_{l}$ because
of additional conditions which are usually related to the linear part
of Eq.\ref{eq:14}. In the case of Eq.\ref{eq:39} would be interesting
whether the positivity principle can be easy satisfied.

\section{{\large{}Matter disappears in zeroth order approximation}}

We assume that the generating vector |V> can be presented as the power
series with respect to the positive or negative powers of the coupling
constant $\lambda$ 

\begin{equation}
|V>\equiv|V(\lambda)>=\sum_{\mu}\lambda^{\mu}|V>^{(\mu)}\label{eq:40}
\end{equation}
 where $\mu$ runs through integers from zero to plus infinity, or
from zero to minus infinity. The first expansion we will use in the
case of Eq.\ref{eq:36}, the second expansion will be used in the
cases of Eq.\ref{eq:37} and Eq.\ref{eq:39}. The vectors $|V>^{(\mu)}$
are called the $\mu$ \textit{approximations}. 

In this Section we use Eq.\ref{eq:36} in which in the 0-approximation
the term with the operator $\hat{N}$ disappears. We may say that
matter, a cause of any interaction, disappears. We get:

\begin{equation}
\left(\hat{I}+\hat{L}_{r}^{-1}\hat{G}\right)|V>^{(0)}=\hat{L}_{r}^{-1}|0>_{info}+\hat{P}_{L_{r}}|V>^{(0)}\label{eq:41}
\end{equation}
For $\hat{G}=0$, and so when external influence, which is produced
by material objects disappears too, zero approximation to the full
generating vector |V> for all n-pi is given by the equation:

\begin{equation}
|V>^{(0)}=\hat{L}_{r}^{-1}|0>_{info}^{(0)}+\hat{P}_{L_{r}}|V>^{(0)}=|0>+\hat{P}_{L_{r}}|V>^{(0)}\label{eq:41-1}
\end{equation}
 The permutation symmetry of n-pi imposes on the generating vector
$|V>^{(0)}$ the condition:

\begin{equation}
|V>^{(0)}=\hat{S}|V>^{(0)}\label{eq:43}
\end{equation}
where the projector $\hat{S}$ is constructed in, e.g., \cite{Hanc 2013}.
A construction in the zeroth approximation of n-pi satisfying these
two equations and the positivity priciple \ref{eq:2-1} one can find
in \cite{Han 2010}. 

Taking into account the permutation symmetry for all n-pi

\begin{equation}
|V>=\hat{S}|V>\label{eq:44}
\end{equation}
 and the \textit{perturbation principle:}

\begin{equation}
|V(0)>=|V>^{(0)}=\hat{S}\hat{P}|V(0)>^{(0)}\label{eq:45}
\end{equation}
where $\hat{P}|V>$ represents the projection occurring in Eq.\ref{eq:36},
one can transform this equation in such a way that every higher approximation
to the generating vector |V>, even in the case of $\hat{G}\neq0$,
is completely expressed by the zeroth order approximation, see also
Secs 6 and 7. So, multiplying Eq.\ref{eq:36} by the projector $\hat{S}$,
we can get: 

\begin{equation}
\left(\hat{I}+\hat{S}\hat{L}_{r}^{-1}\hat{G}+\lambda\hat{S}\hat{L}_{r}^{-1}\hat{N}\right)|V>=\hat{S}\hat{L}_{r}^{-1}|0>_{info}+\hat{S}\hat{P}_{L_{r}}|V>\label{eq:46}
\end{equation}
 Assuming in addition to \ref{eq:45}:

\begin{equation}
\hat{S}\hat{L}_{r}^{-1}|0>_{info}=\hat{S}\hat{L}_{r}^{-1}|0>_{info}^{(0)}=|0>\label{eq:47}
\end{equation}
 we get the following perturbation formulas:

\begin{equation}
\left(\hat{I}+\hat{S}\hat{L}_{r}^{-1}\hat{G}\right)|V>^{(n)}+\left(\hat{S}\hat{L}_{r}^{-1}\hat{N}\right)|V>^{(n-1)}=0\label{eq:48}
\end{equation}
 for n=1,2,...which can be further transform as follows:

\begin{equation}
|V>^{(n)}+\left(\hat{I}+\hat{S}\hat{L}_{r}^{-1}\hat{G}\right)^{(-1)}\left(\hat{S}\hat{L}_{r}^{-1}\hat{N}\right)|V>^{(n-1)}=0\label{eq:49}
\end{equation}

\section{{\large{}Space and time disappear in zeroth approximation}}

We can achieve this situation in the Eq.\ref{eq:15} if you divide
it by $\lambda$, and assuming that in the $\lambda\rightarrow\infty$
solutions are bounded. In this case we get from Eq.\ref{eq:37} and
\ref{eq:44} the following equation: 

\begin{equation}
|V(\infty)>=|V>^{(0)}=\hat{S}\hat{P}_{N_{r}}|V(\infty)>^{(0)}\label{eq:50}
\end{equation}
 which leads, for local interactions, to the product of the Kronecker's
deltas for n-pi. Higher approximations are given by similar formulas
as \ref{eq:49}.

\subsection{{\normalsize{}Higher approximations. Exact solutions?}}

Taking into account the symmetrized Eq.\ref{eq:37}:

\begin{equation}
\left(\lambda^{-1}\hat{S}\hat{N}_{r}^{-1}(\hat{L}+\hat{G})+\hat{I}\right)|V>=\lambda^{-1}\hat{S}\hat{N}_{r}^{-1}|0>_{info}+\hat{S}\hat{P}_{N_{r}}|V>\label{eq:64}
\end{equation}
 we can derive the following formula:

\begin{equation}
|V>\equiv|V(\lambda)>=\left(\lambda^{-1}\hat{S}\hat{N}_{r}^{-1}(\hat{L}+\hat{G})+\hat{I}\right)^{-1}\left(\lambda^{-1}\hat{S}\hat{N}_{r}^{-1}|0>_{info}+\hat{S}\hat{P}_{N_{r}}|V>\right)\label{eq:65}
\end{equation}
 in which the inverse operator is an inverse to the unit and the lower
triangular operator $\hat{S}\hat{N}_{r}^{-1}(\hat{L}+\hat{G})$. It
is interesting to notice that in the extreme case $(\lambda\rightarrow\infty)$
traces of interaction are stuck only in the projector $\hat{P}_{N_{r}}=\hat{I}-\hat{N}_{r}^{-1}\hat{N}_{r}$.
Now, the perturbation principle gives the following identification

\begin{equation}
|V(\infty)>=|V>^{(0)}=\hat{S}\hat{P}_{N_{r}}|V(\infty)>\label{eq:66}
\end{equation}
If the formula \ref{eq:66} satisfies the positivity conditions \ref{eq:2-1},
then the generating vector $|V>$ satisfies these conditions at least
for appropriate small values of $|1/\lambda|$. 

Using the \textbf{perturbation principle}, we identify the zero approximation
with the projection:

\begin{equation}
|V>^{(0)}=\hat{S}\hat{P}_{N_{r}}|V(\infty)>\label{eq:67}
\end{equation}
In this way, for the polynomial interactions, in a finite number of
steps, \textbf{one can receive exact expressions for the n-pi}!! For
the projection properties of operators describing the polynomial interactions
and their inverses, see, e.g., \cite{han 2011,Hanc 2013}.This is
not the case for the theory of perturbations considered in Sec.5,
for which in zero approximation matter disappears. A common feature
of both theory is, however, a similar structure of zero approximations
given respectively by formulas $\ref{eq:45}$ and \ref{eq:66}, see
\cite{Han 2010}.

\section{{\large{}Equations with left invertible interaction; matter, space
and time do not disappear}}

In Eq.\ref{eq:15} three kind of operators appear: $\hat{L},\:\hat{N}\:\hat{G}$
which are related to different physical and mathematical situations.
The operator $\hat{L}$ is related to the linear part of Eq.\ref{eq:14}
which describes \textit{kinematics of the theory}; In other words,
it describe changes, not causes of changes. The causes of changes
are related to the second and third of listed operators even if they
have different characters. $\hat{N}$ describes rather causes (interaction)
among the system elements, e.g., particles. $\hat{G}$ describes causes
(sources of fields or external forces) acting on the system. Looking
at the transformed Eq.\ref{eq:15} given in Sec.4 in three forms:
Eqs \ref{eq:36}, \ref{eq:37} and \ref{eq:39}, and using as in Sec.5
rather unused language, we see that when the matter disappears in
the zeroth approximation then we are getting an expansion of the generating
vector |V> in the positive powers of the coupling constant $\lambda$,
but when the time and space disappear, we are getting an expansion
in the inverse powers of the coupling constatnt. 

In spite of well established habit of assuming that $\hat{N}$ in
the FFS is a right-invertible operator (polynomial interactions),
there are several reasons to give up such assumptions. First of all,
for polynomial interactions the perturbation expansion can be only
used in the two extreme cases: a small and a large absolute value
of the coupling constant $\lambda$. Moreover, even the exact solutions
obtained in Sec.6.1 in the inverse coupling constant $\lambda$, for
which the space and time disapears in the limit $\lambda\rightarrow\infty$,
have no simple physical interpretations.

\subsection{Expansion in the minor (scale) coupling constant $\lambda'$ ; the
left invertible operator $\hat{N}(\lambda')\equiv\hat{N}_{l}(\lambda')$}

The situation may change, when $\hat{N}$ is a left invertble operator
with a left inverse denoted by $\hat{N}_{l}^{-1}$, see Sec.4. In
this case, Eq.\ref{eq:39} can be used:

\begin{equation}
\left(\lambda^{-1}\hat{N}_{l}^{-1}(\lambda')(\hat{L}+\hat{G})+\hat{I}\right)|V>=\lambda^{-1}\hat{N}_{l}^{-1}|0>_{info}=\lambda^{-1}|0>\label{eq:39-1}
\end{equation}
See, however, Subsec.8.2. Taking into account that the operator $\hat{N}=\hat{N}(\lambda')$,
related to the inner and/or external interaction of the system, depends
on the constatnt $\lambda'$ (minor coupling constant) and that the
following expansion is possible:

\begin{equation}
\hat{N}_{l}^{-1}(\lambda')=\hat{N}_{l}^{-1(0)}+\sum_{n=1}\lambda'^{n}\hat{N}_{l}^{-1(n)}\label{eq:54}
\end{equation}
 see Sec.4, Eq.\ref{eq:34} and other, we get, for the zeroth approximation
$|V>^{(0)}$ in the minor coupling constant $\lambda'$, 

\begin{equation}
|V>=|V;\lambda,\lambda'>=|V;\lambda>^{(0)}+\sum_{n=1}\lambda'^{n}|V;\lambda>^{(n)}\label{eq:55}
\end{equation}
the following equation:

\begin{equation}
\left(\lambda^{-1}\hat{N}_{l}^{-1(0)}(\hat{L}+\hat{G})+\hat{I}\right)|V;\lambda>^{(0)}=\lambda^{-1}\hat{N}_{l}^{-1}|0>_{info}^{(0)}=\lambda^{-1}|0>\label{eq:56}
\end{equation}
 In spite of the branching character of this equation, it can be solved
at least for interaction type such as those given by Eq.\ref{eq:33},
for which

\begin{equation}
\hat{N}_{l}^{-1(0)}=\hat{N}_{l}^{-1}(0)=\int d\tilde{y}E(\tilde{y})\hat{\eta}(\tilde{y})+\hat{P}_{0}\label{eq:57}
\end{equation}
 In this case, the zeroth order approximation to Eq.\ref{eq:15} 

\begin{equation}
\left(\hat{L}+\hat{G}+\lambda\hat{N}(0)\right)|V;\lambda>^{(0)}=|0>\label{eq:58}
\end{equation}
is unbranching because

\begin{equation}
\hat{N}^{(0)}\equiv\hat{N}(0)=\int d\tilde{x}\hat{\eta}^{\star}(\tilde{x})H(\tilde{x})+\hat{P}_{0}\label{eq:59}
\end{equation}
 is a lower triangular operator with respect to projectors $\hat{P}_{n}$
projecting on the n-point vectors given by Eqs\ref{eq:7}, see e.g.
\cite{Han 2010}. The Eq.\ref{eq:58} is described by the sum of diagonal
operator $\hat{L}$ and two lower triangular operators $\hat{G}$
and $\lambda\hat{N}(0)$. Eq.\ref{eq:58}, and so the Eq.\ref{eq:56}
can be easy solved in the case of right invertible operator $\hat{L}$
which usually takes place. In this case we can describe its solutions
in the following form:

\begin{equation}
|V;\lambda>^{(0)}=\left[\hat{I}-\hat{S}\hat{L}_{r}^{-1}(\hat{G}+\lambda\hat{N}(0))\right]^{-1}\left(\hat{S}\hat{P}_{L_{r}}|V;\lambda>^{(0)}+|0>\right)\label{eq:60}
\end{equation}
 where the symmetry condition \ref{eq:43} was taken into account.
Here the projector $\hat{P}_{L_{r}}=\hat{I}-\hat{L}_{r}^{-1}\hat{L}$.
Assuming the perturbation principle \ref{eq:45} in the form

\begin{equation}
\hat{S}\hat{P}_{L_{r}}|V;\lambda>^{(0)}=\hat{S}\hat{P}_{L_{r}}|V;0>^{(0)}\label{eq:76-2}
\end{equation}
 we can guarantee that at least for a small external field $\hat{G}$
and a small coupling constant $\lambda$ , the coefficients of the
zeroth order of the generating vector $|V;\lambda>^{(0)}$ satisfies
the positivity condition \ref{eq:2-1}. For a more explicite description
of the above formula, see \cite{Han 2010}. 

For the higher approximations occurring in Eq.\ref{eq:55}, the Eq.\ref{eq:39}
can be used, which we rewrite in the symmetrical form as:

\begin{equation}
\left(\lambda^{-1}\hat{S}\hat{N}_{l}^{-1}(\lambda')(\hat{L}+\hat{G})+\hat{I}\right)|V>=\lambda^{-1}\hat{S}\hat{N}_{l}^{-1}|0>_{info}\label{eq:62}
\end{equation}
 See also Subsec.8.2. It should be noted that to implement the expansion
\ref{eq:55} we do not need to know explicitly the operator $\hat{N}(\lambda')$
. We only need to assume that $\hat{N}(0)$ has the form \ref{eq:59}
which allows us to compute the zero approximation $|V;\lambda>^{(0)}$
given implicitly by formula \ref{eq:60} and we have to know the expicit
form of well defined operator $\hat{N}_{l}^{-1}(\lambda')$, see Sec.4
and Final Remarks.

\section{{\large{}Essentially nonlinear interaction (ENI) }}

Equations for n-pi with polynomial interactions are typically branched
equations, for which the closure problem arises with different ad
hock amputation methods proposed for its solutions. In such situation
the ENI with no closure problems are sought which would approximate
in some way the models with the closure problem, \cite{Han 2008}.
There are also physical reasons to consider such interactions, \cite{Efi 1977}.
Let us now consider Eq.\ref{eq:15} with the left invertible interaction
\ref{eq:45-1}:

\begin{equation}
\left(\hat{L}+\lambda\hat{N}_{l}\frac{\hat{I}}{\hat{I}-\lambda'\hat{N_{r}}}+\hat{G}\right)|V>=|0>_{info}\label{eq:73}
\end{equation}
 The interaction operator $\hat{N}=\hat{N}_{l}\frac{\hat{I}}{\hat{I}-\lambda'\hat{N}}$
appearing in the above equation can be essentially nonlinear (it means
that it corresponds to essentially nonlinear interaction) for two
reasons, namely that $\hat{N}_{l}$ would be essentially nonlinear
and that $\frac{\hat{I}}{\hat{I}-\lambda'\hat{N}_{r}}$ is essentially
nonlinear even if the operator $\hat{N}_{r}$ is related to a polynomial
interaction. Multiplying Eq.\ref{eq:73} by the left inverse operator
\ref{eq:46-1}, we get the equation

\begin{equation}
\left(\left(\hat{I}-\lambda'\hat{N_{r}}\right)\hat{N}_{l}^{-1}(\hat{L}+\hat{G})+\lambda\hat{I}\right)|V>=|0>_{info}\label{eq:74}
\end{equation}
 which is conditionally equivalent to Eq.\ref{eq:73}, see Subsec.8.2.

Assuming as usually that

\begin{equation}
(\hat{L}+\hat{G})\equiv(\hat{L}+\hat{G})_{r}\label{eq:75}
\end{equation}
is a right invertible operator and introducing the operator 

\begin{equation}
\hat{A}_{r}=\lambda'\hat{N_{r}}\hat{N}_{l}^{-1}(\hat{L}+\hat{G})_{r}\label{eq:75-1}
\end{equation}
 one can describe Eq.\ref{eq:74} in an equivalent way:

\begin{equation}
\left(\hat{I}-\hat{S}\hat{A}_{r}^{-1}[\hat{N}_{l}^{-1}(\hat{L}+\hat{G})_{r}+\lambda\hat{I}]\right)|V>=-\hat{S}\hat{A}_{r}^{-1}|0>_{info}+\hat{S}\hat{P}_{A}|V>\label{eq:76-1}
\end{equation}
where a right inverse to the operator $\hat{A}_{r}$ is 

\begin{equation}
\hat{A}_{r}^{-1}=\lambda'^{-1}(\hat{L}+\hat{G})_{r}^{-1}\hat{N}_{l}\hat{N_{r}}^{-1}\label{eq:77}
\end{equation}
 and the projector $\hat{P}_{A}=\hat{I}-\hat{A}_{r}^{-1}\hat{A}_{r}$.
In the Eq.\ref{eq:76-1} we also took into account the permutation
symmetry \ref{eq:44} of solutions.

For $\hat{N}_{l}=\hat{I}$ and the whole set of lower triangular operators
$\hat{N}_{l}$, the Eq.\ref{eq:76-1} is not branching equation in
spite of the fact that Eq. \ref{eq:74} are branched equations. To
fix the arbitrary projection $\hat{P}_{A}|V>$one can use the symmetry
condition and the perturbation principle which relies on assumption
that the arbitrary projection $\hat{P}_{A}|V>$ is constructed by
means of the generating vectors $|V>$solving Eq.\ref{eq:73} with
$\lambda=0$. It means that

\begin{equation}
\hat{S}\hat{P}_{A}|V>=\hat{S}\hat{P}_{\hat{L}+\hat{G}}|V>_{\lambda=0}\label{eq:80-1}
\end{equation}
 We also have to satisfy the positivity conditions for the n-pi which
can be fulfilled by a proper choice of the projection $\hat{P}_{A}|V>|_{\lambda=0}$
and the appropriate choice of parameters like $\lambda$ and $\lambda'$.

\subsection{Another ENI}

That is interaction \ref{eq:47-1}:

\begin{equation}
\hat{N}=\hat{N}'_{l}(\lambda')=\frac{\hat{I}}{\hat{I}-\lambda'\hat{N}_{r}}\hat{N}_{l}\label{eq:81-1}
\end{equation}
 For $\hat{N}_{r}=\hat{N}'_{r}\hat{N}_{l}^{-1}$, we can formally
write:

\begin{equation}
\hat{N}'_{l}(\lambda')=\hat{N}_{l}+\lambda'\hat{N}'_{r}+\lambda'^{2}\hat{N}'_{r}\hat{N}_{l}^{-1}\hat{N}'_{r}+\cdots\label{eq:76}
\end{equation}
If the variables entering the interactions $\hat{N}_{l}$ and $\hat{N}'_{r}$
are independent of each other, it is already the second approximation
to the formula \ref{eq:47-1} in which interaction between them appears.
In this case we have clear separation between ENI and polynomial interactions
exemplified in the second and higher powers of the expansion \ref{eq:76}.

\subsection{A definition of the operator $\frac{\hat{I}}{\hat{I}-\lambda'\hat{N}_{r}}$
in the FFS}

We define this operator as a righ inverse operator to the operator
$(\hat{I}-\lambda'\hat{N}_{r})_{r}$. Here subscript r has to remind
us that the relevant operators are left reversible. This is much weaker
assumtion then an assumption that it is a non singular operator. This
means that in the adopted notation:

\begin{equation}
\frac{\hat{I}}{\hat{I}-\lambda'\hat{N}_{r}}\equiv(\hat{I}-\lambda'\hat{N}_{r})_{r}^{-1}\equiv\hat{Y}\label{eq:83-1}
\end{equation}
So to that specific operator we have to solve the equation: 
\begin{equation}
(\hat{I}-\lambda'\hat{N}_{r})_{r}\hat{Y}=\hat{I}\label{eq:84-1}
\end{equation}
 In fact this equation is like an operator version of equations for
Green's functions. Multiplying this equation by a right inverse to
the operator $\lambda'\hat{N}_{r}$, we get equivalent equation:

\begin{equation}
(\hat{I}-\lambda'^{-1}\hat{N}_{r}^{-1})\hat{Y}=\hat{P}\hat{Y}-\lambda'^{-1}\hat{N}_{r}^{-1}\label{eq:85}
\end{equation}
 from which we get
\begin{equation}
\hat{Y}\equiv\frac{\hat{I}}{\hat{I}-\lambda'\hat{N}_{r}}=(\hat{I}-\lambda'^{-1}\hat{N}_{r}^{-1})^{-1}\left(\hat{P}\hat{Y}-\lambda'^{-1}\hat{N}_{r}^{-1}\right)\label{eq:86}
\end{equation}
 Here the projector $\hat{P}=\hat{I}-\hat{Q}$ where the projector
$\hat{Q}=\hat{N}_{r}^{-1}\hat{N}_{r}$ and the projection $\hat{P}\hat{Y}$
describes an arbitrary term which has to be specified by means of
additional conditions to Eq.\ref{eq:85}, see \cite{Przew 1988}. 

For a local theories, we should rather choose the local operators
$\hat{N}_{r}^{-1}$. The arbitrary projection $\hat{P}\hat{Y}$ can
be restricted by the positivity properties of n-pi when at least all
arbitrary pairs of variables $\tilde{x}_{j};j=1,...,n$ are equal. 

One can show that for

\begin{equation}
\hat{P}\hat{Y}=0\label{eq:87}
\end{equation}
 
\begin{equation}
\hat{Y}=\hat{Y}\hat{Q}\label{eq:88}
\end{equation}
 and in this case the natural property: 

\begin{equation}
\hat{N}_{r}\frac{\hat{I}}{\hat{I}-\lambda'\hat{N}_{r}}=\frac{\hat{I}}{\hat{I}-\lambda'\hat{N}_{r}}\hat{N}_{r}\label{eq:89}
\end{equation}
takes place, see also \cite{Hanc 2013}. 

PROBLEM:

It is not clear if for the choice \ref{eq:87} the positivity properties
of n-pi are satisfied, for all equal pairs of 'points' $\tilde{x}_{j};j=1,...,n$.
In other words, this choice can be treated as the first approximation
to a proper definition of the operator $\frac{\hat{I}}{\hat{I}-\lambda'\hat{N}_{r}}$.
At this point we would like to notice that a 'proper definition' of
the operator depends on the context which in our case is Eq.\ref{eq:15}.
At this point we would also like to notice that in the situation of
such a dramatice freedom of solutions in the FFS, the positivity condition
of n-pi is a very important tool ensuring a proper relationship of
equations \ref{eq:15} with the original equations \ref{eq:14}. 

COMMENT1

The choice \ref{eq:87} in the case of polynomial operators $\hat{N}_{r}$
leads to lower triangular operators in Eq.\ref{eq:15} which usually
do not appear in such equations, see e.g. \cite{han 2011}.

COMMENT2

A proper definition of the ENI $\frac{\hat{I}}{\hat{I}-\lambda'\hat{N}_{r}}$
is also important for such a reason that with its help we can constract
other operators, see Hilbert transforms and \cite{Hanc 2013,Han 2011'}. 

COMENT3

However, the positivity condition \ref{eq:2-1} is required if the
values of the field $\varphi(\tilde{x})$ are numbers. If this is
not the case and values of the field are operators, then this condition
need not occur. Because in both cases we are dealing with the same
equations in terms of the form, we have here the same situation as
in classical and quantum mechanics, where the same equations in terms
of the form describe physical systems differing only in the micro
and macro scale. See planetary Bohr atomic model. I would like to
stress that in the case of all functions $\beta$ equal to each other
and describing initial and boundary conditions for Eq.\ref{eq:14}
the lack of positivity conditions for n-pi can be most likely interpreted
as a lack of commutativity among values of the field $\varphi(\tilde{x})$
at differen points $\tilde{x}$. In such a case do we need to know
commutation relations of the implicitly appearing operators? Perhaps
- no - if we are able to interpret n-pi not satisfying the positivity
principle.

\section{{\large{}Overdetermined equations}}

An equation is overdetermined if its multiplication by the projector
$\hat{Q}$ does not change its set of possible physical solutions.
In the FFS it is the case when physical solutions have certain symmetries.

Let us multiply Eq.\ref{eq:15} 

\[
\left(\hat{L}+\lambda\hat{N}_{l}+\hat{G}\right)|V>=|0>_{info}
\]
 by a left invers operator $\hat{N}_{l}^{-1}$ . We get

\begin{equation}
\left(\hat{N}_{l}^{-1}(\hat{L}+\hat{G})+\lambda\hat{I}\right)|V>=\hat{N}_{l}^{-1}|0>_{info}=|0>_{info}\label{eq:80}
\end{equation}
 Multiplying Eq.\ref{eq:80} by $\hat{N}_{l}$, we get 

\begin{equation}
\left(\hat{Q}(\hat{L}+\hat{G})+\lambda\hat{N}_{l}\right)|V>=|0>_{info}\label{eq:81}
\end{equation}
 Assuming that $\hat{N}_{l}=\hat{Q}\hat{N}_{l}$ with the projector
$\hat{Q}\equiv\hat{N}_{l}\hat{N}_{l}^{-1}$ and that $|0>_{info}=\hat{Q}|0>_{info}$,
we see that Eq.\ref{eq:81} is equivalent to Eq.\ref{eq:80} only
if this equation is overdetermined in such a way that the multiplication
by the projector $\hat{Q}$ does not change its set of possible physical
solutions. In the FFS it is the case especially when the physical
solutions have certain symmetries, for example: the permutations or
Poincare symmetry.

\section{{\large{}Complete and closed sets of n-pi}}

The set of n-pi is \textit{complete}, if we are able to derive equations
for them by means of which one can construct these n-pi. A finite
set of n-pi is \textit{closed} if is also complete. Equations as \ref{eq:15},
containing an upper triangular operator, usually produce unclosed
sets of n-pi. Then we say that the closure problem arises. The closure
problem is \textit{apparent} if after appropriate transformation of
a considered equation this problem disappears.

\section{{\large{}A construction to Eq.\ref{eq:14} the zeroth order n-pi
satisfying the positivity conditions }\ref{eq:2-1} }

In this case Eq.\ref{eq:14} looks as follows

\begin{equation}
L[\tilde{x};\varphi]+G(\tilde{x})=0\label{eq:96}
\end{equation}
 The general solution to this equation is represented in the form

\begin{equation}
\varphi[\tilde{x};\alpha]=\varphi_{0}[\tilde{x};\alpha]-\int d\tilde{u}L_{r}^{-1}(\tilde{x},\tilde{u})G(\tilde{u})=\int d\tilde{u}\, P_{0}(\tilde{x},\tilde{u})\alpha(\tilde{u})-\int d\tilde{u}L_{r}^{-1}(\tilde{x},\tilde{u})G(\tilde{u})\label{eq:97}
\end{equation}
 where the function $P_{0}$ is a kernel of projector projecting on
the free solutions to Eq.\ref{eq:96} $(G=0)$ and $L_{r}^{-1}$ is
a right inverse to the linear operator $L$. In fact, in the symbolical
notation we have

\begin{equation}
P_{0}=I-L_{r}^{-1}L\label{eq:98}
\end{equation}
 and

\begin{equation}
\varphi[\tilde{x};\alpha]=P_{0}[\tilde{x};\alpha]-L_{r}^{-1}G[\tilde{x}]\label{eq:99}
\end{equation}
 c.f \ref{eq:97}. Here, all operators without Asian hat, \textasciicircum{}
, act in the function space, which are defined on the same manifold
which are defined the fields $\varphi(\tilde{x})\equiv\varphi[\tilde{x};\alpha]$. 

To generate n-pi \ref{eq:2} we introduce the generating functional
(g.f.)

\begin{equation}
V[j]=\int\delta\alpha\, e^{ij\cdot\varphi[\alpha]}P[\alpha]\label{eq:100}
\end{equation}
 from which 

\begin{equation}
\frac{\delta^{n}}{\delta j(\tilde{x}_{1})\cdots\delta j(\tilde{x}_{n})}V[j]_{j=0}=i^{n}<\varphi(\tilde{x}_{1})...\varphi(\tilde{x}_{n});0>\label{eq:101}
\end{equation}
 where zero in the r.h.s of the above formula means that generating
n-pi are permutation symmetrical. In the exponent of formula \ref{eq:100}
$\varphi[\alpha]\cdot j\equiv\int d\tilde{x}\varphi[\tilde{x};\alpha]j(\tilde{x})$,
where $\varphi[\tilde{x};\alpha]$ is a general solution of Eq.\ref{eq:14}.
In the case of Eq.\ref{eq:96}, The general solution has the form
\ref{eq:97}, so we can write the formula \ref{eq:100} as 

\begin{equation}
V[j]=e^{iL_{r}^{-1}G\cdot j}\int\delta\alpha\, e^{iP_{0}\alpha\cdot j}P[\alpha]\label{eq:102}
\end{equation}
 with the help of which, by the n-th order functional differentiations
with respect to function $j(\tilde{x})$, one can generate the permutation
symmeric n-pi, for a given probability distribution $P[\alpha]$.
Of course, to get a final answer we have yet to execute the functional
integration $\int\delta\alpha\cdots$, which only can be done, for
a Gaussian type of probability distribution $P[\alpha]$, cf. e.g.
\cite{rzew 1969}. In other cases, one can use equations as Eq.\ref{eq:15}
or equations presented in the paper in which we do not need to know
the general solutions to Eq.\ref{eq:14}. For the Gaussian probability
density $P=exp(-1/2\cdot\alpha^{2})$ we get 

\begin{equation}
V[j]=e^{iL_{r}^{-1}G\cdot j}e^{-\frac{1}{2}(P_{0}j)^{2}}\label{eq:103}
\end{equation}
 It is also well known that for certain averages we do not need to
know general solutions to Eq.\ref{eq:14}. This happens when the probability
distribution $P$ is related to a certain integral of motion to Eq.\ref{eq:14},
for example the energy integral $H[\alpha(t,\cdot)]$. Of course,
it is not the case of theory with $G\neq0$.

\section{S{\large{}chwinger equations and some surprising parallels}}

If the system is complicated and/or is very sensitive to small changes
in initial and/or boujndary conditions, then it does not matter any
detailed his description, and begin to be of interest averaged descriptions
as n-pi. Then we replace Eqs \ref{eq:14}, which provide a detailed
description of the system, by Eqs \ref{eq:15} for averaged or smoothed
quantities called n-pi in which the interactions and kinematics of
the system are described by the operators in the FFS. Consider again
g.f. \ref{eq:100}, but this time with the probability distribution
P given by the energy integral $H$:

\begin{equation}
V[j]=\int\delta\alpha\, e^{ij\cdot\varphi[\alpha(t^{0},\cdot)]}e^{-\beta\int dt\, H[\alpha(t^{0},\cdot)]}=\int\delta\alpha\, e^{ij\cdot\varphi[\alpha(t^{0},\cdot)]}e^{-\beta\int dt\, H[\varphi[\alpha(t^{0},\cdot)]]}\label{eq:104}
\end{equation}
Here $\beta$ is a dimensional constant. Denoting the invertible transformation
$\alpha\rightarrow\zeta$ : $\zeta(\tilde{x})=\varphi[\tilde{x};\alpha]$,
and assuming that 

\[
(i)\;\delta\zeta=\delta\alpha
\]
(cf. Liouville's theorem) we can write the above g.f. as follows:

\begin{equation}
V[j]=\int\delta\zeta e^{ij\cdot\zeta}e^{-\beta\int dt\, H[\zeta(t,\cdot)]}\label{eq:105}
\end{equation}
 Additionally assuming the translational invariance of measure: 

\[
(ii)\;\delta\zeta=\delta(\zeta+\gamma)
\]
for an arbitrary function $\gamma$, we can derive the Schwinger equations,
see \cite{rzew 1969}, for the g.f. $V[j]$ in the case of averagings
given by the probability distribution

\begin{equation}
P[\zeta]=e^{-\beta\int dt\, H[\zeta(t,\cdot)]}\label{eq:106}
\end{equation}
 As you see, the above functional integral can be constructed without
using any solutions to Eq.\ref{eq:14}. It satisfies Schwinger equation
which we now derive. To do this we have to do a third assumption that
the functional derivative commutes with the functional integration:

\[
(iii)\;\left[\int,\delta/\delta\gamma(\tilde{x})\right]=0
\]
for an arbitrary $\tilde{x}$ and $\gamma(\tilde{x})=0$. We start
from

\begin{equation}
V[j]=\int\delta\zeta e^{ij\cdot(\zeta+\gamma)}e^{-\beta\int dt\, H[\zeta(t,\cdot)+\gamma(t,\cdot)]}\label{eq:107}
\end{equation}
 which we differentate with respect to $\gamma$ at $\gamma=0$. In
result we get:

\begin{equation}
\int\delta\zeta\cdot\{ij(\tilde{x})-\beta H'[\zeta;\tilde{x}]\}e^{ij\cdot\zeta}e^{-\beta\int dt\, H[\zeta(t,\cdot)]}=0\label{eq:108}
\end{equation}
 with $H'[\zeta;\tilde{x}]\equiv\delta/\delta\gamma(\tilde{x})\int dt\, H[\zeta(t,\cdot)+\gamma(t,\cdot)]_{\gamma=0}$.
Using the formula for integration by parts, the last equation can
be rewritten as

\begin{equation}
\left\{ ij(\tilde{x})-\beta H'[\tilde{x};-i\delta/\delta j]\right\} V[j]=0\label{eq:109}
\end{equation}
 This is the Schwinger equation, which differs from Eq.\ref{eq:15}
by the first term corresponding to the lower triangular operator,
see \cite{han 2011}. The advantages of using this equation is based
on the fact that obtained in this way n-pi satisfy the positivity
conditions, which is visible from the formula \ref{eq:105} of the
generating functional V. It is also interesting that derived in this
way n-pi are multi-time. To these multi-time n-pi correspond averages
with the probability distribution given by the energy integral. To
construct in this way the n-pi any exact solutions to Eq.\ref{eq:14}
are not used at all. Moreover, even if assumptions $(i-iii)$ are
not correct, obtained in this way n-pi can be interpretowanr as averages
describing the systems in whose the total energy is not specific.
Such situations could correspond to action on the systems unknown
external forces changing randomly their energy. 

What is interesting here is that the transition in \ref{eq:105} to
the purely imaginary time corresponds to the transition to quantum
field theory (Green's functions), in which the phenomena are described
in which one member of each pair of coupled variables (position or
momentum) are not specified. In some way one can say that a property
of the impossibility of getting full information about every pair
of conjugate variables, in quantum field theory (QFT), is expressed
by introduction of the imaginary time. See also \cite{Vasy 2015},
where the imaginary time is also interpreted as an impossibility of
happining certain events in QM. See also \cite{Han 2015}. However,
the introduction of imaginary time leads to the appearance of the
Fresnel types integrals, that require appropriate modifications of
integrals, see \cite{rzew 1969}. 

In QFT, the higher n-pi (Green's functions) are related to a creation
and annihilation of more particles in the scattering experiments.
In classical mechanics or field theory, the higher n-pi are related
to extra information about the system. Is it possible that information
also had a dual nature?

To describe the Schwinger Eq.\ref{eq:109} in the FFS, see, e.g.,
\cite{han 2011}(?).

\section{{\large{}Remarks}}

Let us also notice that in the free Fock space (FFS) even in the simple
mathematical model with an essentially nonlinear (nonpolynomial) interaction
(ENI), Eq.\ref{eq:33}, which is represented by the left invertible
operator, successive approximations to it, given by formulas such
as Eq.\ref{eq:31}, are described by right invertible operators. In
fact only the zeroth approximation describing external fields is a
lower triangular operator, see Eq.\ref{eq:18}. Electromagnetic, strong,
and weak interactions are polynomial type. It is not out of the question
that the essentially nonlinear left invertible operator $\hat{N}$,
which generates in the expansion with respect to the parameter $\lambda'$
the right invertible operators, is the prototype of an operator describing
theory of everything containing in itself also the essentially nonlinear
gravitation theory? 

We would like to underline once again the importance of the auxiliary
parameter $\lambda'$whose introduction makes possible to execute
the perturbation calculations \textbf{without dramatice change of
the whole theory at the zeroth order approximation}: ``no matter,
or no space-time assumptions'', see Secs 5 and 6. In fact we do such
change only for the appropriate projection, see \ref{eq:80-1}. The
introduction of this parameter via ENI allows for multiscale description
of the system in which these basic concepts do not disapear at any
scale. Furthermore, since in some cases the introduction of essentially
nonlinear interaction (ENI) leads to close branching equations, \textbf{it
is a rare instance when it seems that the complexity issue is preferred}.
See also \cite{Hanc 2010'}where it is shown that very complicated
Lagrangians with ENI lead to closed equations. 

In fact, we think that this is a \textbf{relict} of the period, when
the stable solutions of nonlinear equations \ref{eq:14} were the
main instrument for learning about Nature. At presence, when the focus
of the study is shifted to the equations with unstable solutions which
are described by the \textbf{linear equations} discussed inter alia
in the submitted paper, the precise or approximate solutions should
be constructed on the basis of an one-side invertability of the operators
and their corresponding definitions in the FFS, which occur at these
equations. Of course, we can not forget about an influence of a concrete
choice of averaging process on the n-pi. When we claiming that the
polynomial approximation for ENI is possible, it means that the field
$|\varphi|$ can not to be too large which limits the choice of the
functional $P[\gamma]$ in the formula \ref{eq:6-1}. 

I believe that the introduction of general n-pi \ref{eq:6-1} allows
for more complete extraction of the physical content of the theory,
but also allows a new look at such issues as the perturbation theory,
the choice of additional conditions or definitions of operator-valued
functions. This belief stems from the fact that every general n-pi
depends only on n independent $\beta$ functions, see \ref{eq:6-1},
and this leads to significant restrictictions for the structure of
equations and solutions considered. From this point of view, for example,
the definition (\ref{eq:86},\ref{eq:87}) of the operator $\frac{\hat{I}}{\hat{I}-\lambda'\hat{N}_{r}}$
seems to be appropriate. On the other hand, it requires reflection
the fact that the common source of this or that definition is simple
function $\frac{1}{1-x}$. Could it was a one more sign of reflection
of the simple fundamental laws underlying the Universe? The fact that
for a finite $x$ the function $\frac{1}{1-x}$ is singular may means
that something peculiar is happening in a theory with such singularity
as, e.g., the creation of particles or other things.

\end{document}